\documentstyle[multicol,aps]{revtex}
\begin{document}
\newcommand{\la} {\langle}
\newcommand{\ra} {\rangle}

 \title{Intermittency corrections to the mean square particle acceleration in high Reynolds number turbulence}

\author{Mark Nelkin$^{1}$ and Shiyi Chen$^{2}$ }

\maketitle

\begin{abstract}
The mean square particle acceleration in high Reynolds number turbulence is dominated by the mean square pressure gradient. Recent experiments by Voth et al [Phys. Fluids {\bf 10}, 2268 (1998)] indicate that this quantity, when normalized by the 1941 Kolmogorov value, ${\epsilon}^{3/2}{\nu}^{-1/2}$, is independent of Reynolds number at high Reynolds number.  This is to be contrasted with direct numerical simulations of Vedula and Yeung [Phys. Fluids {\bf 11}, 1208, (1999)] which show a strong increase of the same quantity with increasing Reynolds number.  In this paper we suggest that there is no inherent conflict between these two results.  A large part of the increase seen in DNS is shown to be associated with finite Reynolds number corrections within the quasi-Gaussian approximation.  The remaining intermittency corrections are relatively slowly increasing with Reynolds number, and are very sensitive to subtle cancellations among the longitudinal and transverse contributions to the mean square pressure gradient.  Other possible theoretical subtleties are also briefly discussed.

\end{abstract}

\bigskip

\noindent $^{1}$Physics Department, New York University, New York, NY 10003,\\
and Levich Institute, CCNY, New York, NY 10031 USA.\\
Electronic address: Mark.Nelkin@nyu.edu\\
$^{2}$Center for Nonlinear Studies,
Los Alamos National Laboratory, Los Alamos, NM 87545 USA.\\
Electronic address: syc@cnls.lanl.gov

\begin{multicols}{2}
\narrowtext



There has been considerable recent interest in the study of fluid particle acceleration in turbulent flows.  A natural focus for study is the scaled mean square fluid particle acceleration

\begin{equation}
\beta={\nu}^{1/2}{\epsilon}^{-3/2} \langle a^2 \rangle,
\label{accel}
\end{equation}
where ${\bf a}=d{\bf v}/dt$ is the fluid particle acceleration, $\nu$ is the kinematic viscosity of the fluid, and $\epsilon$ is the average rate of energy dissipation per unit mass.  In the 1941 Kolmogorov theory, $\beta$ is independent of Reynolds number.

Direct numerical simulations of isotropic turbulence by Vedula and Yeung  \cite{yeung98} and by Gotoh and Rogallo \cite{gotoh98} show that the scaled acceleration (\ref{accel}) increases strongly with Reynolds number, approximately as $R_{\lambda}^{1/2}$, where $R_{\lambda}$ is the Taylor microscale Reynolds number.  These DNS also give the expected result that the dominant contribution to $\beta$ comes from the mean square pressure gradient. The viscous contribution is no more than a few percent, and will not be considered further in this paper.  The DNS are currently limited to a maximum $R_{\lambda}$ of 235.  By contrast, a recent high resolution measurement of fluid particle acceleration by Voth et al \cite{voth98} suggests that $\beta$ is nearly independent of Reynolds number in a range of $R_{\lambda}$ between 1000 and 2000.  This result is for the flow between counter-rotating disks. The quantity $\beta$ used in this paper is a factor of $3$ times the scaled acceleration variance $a_0$ used in \cite{voth98} and \cite{yeung98}.  These references look at a single Cartesian component, and we look at the sum over Cartesian components.

We start from an exact result for the mean square pressure gradient for isotropic turbulence due to Hill and Wilczak \cite{hill95},  

\begin{equation}
\langle ({\bf\nabla}p)^2 \rangle = 4 \int_0^{\infty} r^{-3} [L(r)+T(r)-6M(r)] dr,
\label{hill}
\end{equation}
where we have taken the density $\rho=1$ with no loss of generality.  In (\ref{hill}) the quantities $L(r)$, $T(r)$, and $M(r)$ are the three independent fourth order velocity structure functions. We refer to these as the longitudinal $L(r)$, transverse $T(r)$, and mixed $M(r)$, and they are given by

\begin{equation}
L(r) = \la \Delta u^4(r) \ra, \;  T(r) =\la \Delta v^4(r) \ra, \;
 M(r) = \la \Delta u^2(r) \Delta v^2(r) \ra .
\label{struct}
\end{equation}

In a recent letter \cite{nel98}, we have analyzed a similar expression for the pressure structure function $D_p(r)$ in the inertial range, and we found it to be very sensitive to strong cancellations among the positive terms proportional to $L(r)$ and $T(r)$ and the negative term proportional to $M(r)$.  We expect a comparable degree of cancellation here for the mean square pressure gradient. To estimate this cancellation, it is useful to introduce the quasi-Gaussian approximation, where

\begin{equation}
L(r)  = 3 [D_{LL}(r)]^2, T(r)=3 [D_{NN}(r)]^2,
 M(r) = D_{LL}(r) D_{NN}(r).
\label{gauss}
\end{equation}
In (\ref{gauss}), $D_{LL}(r)$ is the second order longitudinal velocity structure function, and $D_{NN}(r)$ is the second order transverse velocity structure function.  For isotropic turbulence, these two quantities are related by

\begin{equation}
D_{NN}(r) = D_{LL}(r) + (r/2) [dD_{LL}(r)/dr].
\label{isotropy}
\end{equation}
If  (\ref{gauss}) and (\ref{isotropy}) are substituted into (\ref{hill}), we obtain

\begin{equation}
\langle ({\bf\nabla}p)^2 \rangle = 3 \int_0^{\infty} r^{-1} [dD_{LL}(r)/dr]^2 dr,
\label{batch}
\end{equation}
which was derived by Obukhov in 1949 \cite{obukhov49}
and by Batchelor in 1951 \cite{batchelor51}. To evaluate (\ref{batch}) at high Reynolds numbers, Batchelor introduced an interpolation formula for the longitudinal structure function, which remains an accurate and useful approximation today.  He suggested that

\begin{equation}
D_{LL}(r)=(1/15){\epsilon}^{2/3} {\eta}^{-2/3} x^2 (1+\alpha x^2)^{-2/3},
\label{interpolate}
\end{equation}
where
$$x=r/\eta,$$
 $\epsilon$ is the average rate of energy dissipation per unit mass, $\eta={\nu}^{3/4}{\epsilon}^{-1/4}$, is the Kolmogorov dissipation length scale, and $\nu$ is the kinematic viscosity of the fluid.  The adjustable parameter $\alpha$ is taken as $\alpha=0.006455$, as discussed elsewhere \cite{review}.  Substituting (\ref{interpolate}) into (\ref{batch}), the integral in (\ref{batch}) is easily evaluated to give

\begin{equation}
{\beta}^{QG}={\epsilon}^{-3/2} {\nu}^{1/2}\langle ({\bf\nabla}p)^2 \rangle =(4/175\alpha)\approx3.5
\label{refvalue}
\end{equation}
A survey of earlier results for the value of ${\beta}^{QG}$ is given by Kaneda \cite{kaneda93}.


At low Reynolds number, there are two distinct corrections to (\ref{refvalue}) to be expected.  Within the quasi-Gaussian approximation, there will be a finite Reynolds number correction due to the cutoff of (\ref{batch}) at large $r$.  This occurs approximately at the integral length scale $L$. This leads to a viscosity independent subtraction from the mean square pressure gradient in (\ref{batch}).  Since the high Reynolds number limit of $\langle ({\bf\nabla}p)^2 \rangle$ is proportional to ${\nu}^{-1/2}\sim R_{\lambda}$, the quasi-Gaussian value of $\beta$ should have a Reynolds number dependence,

\begin{equation}
{\beta}^{QG}=B(1- C /{R_{\lambda}})
\label{lowre}
\end{equation}
It is difficult to estimate the coefficient $C$ from theory, but fortunately the DNS can separately evaluate the full value of $\beta$ and its approximate value from the quasi-Gaussian approximation.  These values, along with the ratio of the full to quasi-Gaussian contribution are given in Table 1.  The full value is from \cite{yeung98}, and the quasi-Gaussian value has been computed from the same data by Vedula and Yeung, and sent to us privately.  The full value is three times the value reported in table 1 of \cite{yeung98} since it is the sum over all three Cartesian components which is used here.

The values of ${\beta}^{QG}$ are well fit by (\ref{lowre}) with $B=3.75$ and $C=14.7$.  The value of $B$ is reasonably close to the value of $B=3.5$ in (\ref{refvalue}).  We have no interpretation for the numerical value of $C$.  Since the large $r$ cutoff is in the energy containing range, this value should depend on the large scale flow conditions, and is not expected to be universal.  

The ratio of the full value to the quasi-Gaussian value is determined by non-Gaussian corrections in the dissipation range of scales $r$.  These corrections are well known to increase as the scale size $r$ decreases.  In the inertial range, they are characterized by the well-known anomalous scaling exponents for the fourth order structure functions.  The situation in the dissipation range is more complicated and less well characterized, but the dominant physical phenomenon is still the increasing intermittency with decreasing scale size.   We thus refer to the corrections to the quasi-Gaussian value as intermittency corrections.  In Table 1, we see that ${\beta}^{QG}$ increases by a factor of  $1.58$ from the smallest to the largest Reynolds numbers in the DNS, and that the intermittency correction increases by a factor $1.66$ in the same range.  For higher Reynolds numbers, however, (\ref{lowre}) indicates that 
${\beta}^{QG}$ is essentially constant.  The Reynolds number dependence of the intermittency correction is difficult to estimate since the cancellation of the terms in (\ref{hill}) can be a sensitive function of Reynolds number.  From Table 1, these corrections appear to increase approximately as

\begin{equation}
\beta/{\beta}^{QG}= Const. R_{\lambda}^{0.23},
\end{equation}
but the power law fit is not very accurate, showing marginal signs of flattening at higher $R_{\lambda}$.


The experiment by Voth et al \cite{voth98} is in the flow between counter-rotating disks in a range of $R_{\lambda}$ from $1000$ to $2000$.  The measured value of $\beta$ in this range is independent of $R_{\lambda}$ and is $21\pm9$.  Most of the uncertainty is in the absolute value.  The Reynolds number dependence is known fairly accurately.  If we extrapolate the intermittency correction from the DNS of Vedula and Yeung, we obtain a value of $\beta=14.0$ at $R_{\lambda}=1000$ and $\beta=16.4$ at $R_{\lambda}=2000$.  We do not know if the experiment is accurate enough to exclude this weak increase in this Reynolds number range, but the absolute value is large enough that substantial corrections to the quasi-Gaussian approximation are surely present.

Now let us examine some of the subtle effects which can cause a complicated Reynolds number dependence of the intermittency corrections.  First look at (\ref{hill}) in the quasi-Gaussian approximation.    All of the integrands show a broad peak in the neighborhood of $r=10\eta$.   Substituting (\ref{gauss}), (\ref{isotropy}) and (\ref{batch}) into (\ref{hill}), all of the integrals can be done analytically to give

\begin{equation}
{\beta}^{QG}=(12/225\alpha)(1.5000+3.4286-4.5000)=(4/175\alpha)\approx3.5,
\label{hillgauss}
\end{equation}
where the first positive term comes from $L(r)$, the second positive term from $T(r)$, and the negative term from $M(r)$.  The cancellation among the three terms is substantial.  If the intermittency corrections to each term have slightly different Reynolds number dependence, this slight difference can be magnified in an unpredictable way in the net value of $\beta$.  Further, the Hill-Wilczak formula (\ref{hill}) is for isotropic turbulence, and is sensitive to corrections due to anisotropy.  The experiment is not fully isotropic even at dissipation scales.  Corrections due to anisotropy can be substantial, and nothing is known about their Reynolds number dependence.

Neglecting such possible differences, it is still a subtle and difficult question to estimate the Reynolds number dependence of the intermittency correction to the longitudinal contribution.  This is governed, to a good approximation by $F(10\eta)/3$, where $F(r)$ is the longitudinal flatness factor

\begin{equation}
F(r)=\la \Delta u^4(r) \ra/\la \Delta u^2(r) \ra^2,
\end{equation}
which is $3$ for a Gaussian distribution.  The crossover of this flatness from the inertial to the dissipation range is quite complicated because of the multifractal distribution of dissipation length scales.  A specific model of this multifractal crossover has been calculated by Eggers and Wang \cite{wang98}.  Working with the input data to Fig. 6 of that paper, we have calculated the correction factor $F(10\eta)/3$ versus Reynolds number in the Reynolds number range of \cite{voth98}.  It increases somewhat more strongly than the extrapolation from the DNS, and thus does not by itself explain the observed plateau in the scaled acceleration variance.

We note that the longitudinal velocity derivative flatness in the flow between counter-rotating disks has been observed \cite{belin97} to have a plateau in its Reynolds number dependence in the same range of Reynolds numbers studied by Voth et al.  The origin of this plateau is not understood.  It is not known if it is a general feature of turbulence at these Reynolds numbers, or if it is a special feature of the flow between counter-rotating disks. The present study started with the idea that these two experiments might be  related, but we have not found any simple connection between them.


To summarize, there is no essential disagreement between the low Reynolds number DNS and the high Reynolds number experiment.  In the DNS, a large part of the increase in the scaled mean square pressure gradient $\beta$ arises from finite Reynolds number corrections within the quasi-Gaussian approximation, and these corrections are obtained directly from the DNS.  These corrections are negligible for the high Reynolds number experiment.  The remaining intermittency corrections are increasing rather slowly with Reynolds number, and it is very difficult to estimate how they will extrapolate to the Reynolds number range of the experiment.  The experimental value of $\beta=21\pm9$ has considerable uncertainty in its absolute value, but it is almost surely much larger than the quasi-Gaussian value, which can be accurately calculated as ${\beta}^{QG}\approx 3.5$.  The approximate Reynolds number independence observed for $\beta$ does not mean that the experiment is consistent with 1941 Kolmogorov scaling, but only that the intermittency corrections, though large, are not varying rapidly with Reynolds number in the range of the experiments.


We would like to thank Eberhard Bodenschatz, Patrick Tabeling, P.K. Yeung, Greg Voth and Victor Yakhot for useful conversations concerning  a preliminary version of this paper.  We would also like to thank Prakash Vedula and P.K. Yeung for calculating the quasi-Gaussian contribution from their DNS, and for sending the results to us before publication, and Jane Wang for giving us the input data for Figure 6 in \cite{wang98}. Finally we would like to thank K. R. Sreenivasan for pointing out the possible importance of anisotropy in interpreting the high Reynolds number experiment.

\bigskip

\begin{table}

\begin{center}
\begin{tabular}{cccc}
$R_{\lambda}$ & $\beta$ & ${\beta}^{QG}$ & Ratio \\
\hline
38 & 3.60 & 2.25 &1.60 \\
90 & 6.81 & 3.19 & 2.14 \\
140 & 8.16 & 3.37 & 2.42 \\
235 & 9.39 & 3.55 & 2.65
\end{tabular}
\caption{Scaled mean square pressure gradient and its
quasi-Gaussian value from the DNS of Vedula and Yeung.}
\label{table1}
\end{center}
\end{table}

\end{multicols}

\end{document}